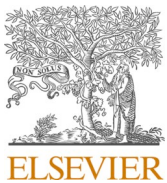
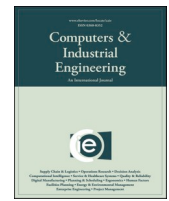

# Brand Network Booster: A new system for improving brand connectivity

Jacopo Cancellieri [1], Walter Didimo [1], Andrea Fronzetti Colladon [1,*], Fabrizio Montecchiani [1], Roberto Vestrelli [1]

*Department of Engineering, University of Perugia, Via G. Duranti 93, 06125 Perugia, Italy*

ARTICLE INFO

*Keywords:*
Semantic network analysis
Text mining
Brand connectivity
Maximum betweenness improvement
Decision support system

ABSTRACT

This paper presents a new decision support system offered for an in-depth analysis of semantic networks, which can provide insights for a better exploration of a brand's image and the improvement of its connectivity. In terms of network analysis, we show that this goal is achieved by solving an extended version of the Maximum Betweenness Improvement problem, which includes the possibility of considering adversarial nodes, constrained budgets, and weighted networks – where connectivity improvement can be obtained by adding links or increasing the weight of existing connections. Our contribution includes a new algorithmic framework and the integration of this framework into a software system called Brand Network Booster (BNB), which supports brand connectivity evaluation and improvement. We present this new system together with three case studies, and we also discuss its performance. Our tool and approach are valuable to both network scholars and in facilitating strategic decision-making processes for marketing and communication managers across various sectors, be it public or private.

## 1. Introduction

A system that can automatically evaluate brand equity by analyzing a company's stakeholders' discourse has become essential in the era of big data. Indeed, textual data is ubiquitous and often available online from freely accessible sources. For example, the relevance and popularity of brands can be assessed by analyzing consumers' posts on social media platforms (Gloor et al., 2009). In the era of social media, actively monitoring online conversations about brands is crucial for effective brand management. These discussions involve individuals who are genuinely invested in brands, providing real-time and dynamic feedback (Rust et al., 2021). Similarly, the media's perspective and the importance they attribute to a specific brand can be evaluated by looking at online news, which can ultimately prime readers and impact their brand awareness and related behaviors (Fronzetti Colladon, 2020; Greco & Polli, 2020; Oliver et al., 2020). In addition, analyzing textual data containing stakeholders' opinions and concerns regarding various issues has proven to be valuable in quantifying the risks to which companies are exposed, often in near real-time (Hassan et al., 2023; Sautner et al., 2023). The analysis of (online) textual data has the advantage of considering the expressions of a high number of stakeholders at a low cost while also reducing some of the biases that can arise when administering brand equity surveys (Bhoi et al., 2022; Grandcolas et al., 2003; Olson, 2006; Wang et al., 2023; Warner, 1965; Zhu et al., 2023). Traditional survey methods often suffer from nonresponse and social desirability biases, which can skew results (Fronzetti Colladon, 2018; Keeter et al., 2006). In contrast, analyzing spontaneous online discussions provides more authentic insights into stakeholders' true opinions and attitudes (Fronzetti Colladon, 2018; Rust et al., 2021; Zhang & Moe, 2021).

In this paper, we consider the conceptualization of brand importance that has been proposed by Fronzetti Colladon (2018), who introduced the Semantic Brand Score (SBS) composite indicator and a system for brand and business intelligence (Grippa & Fronzetti Colladon, 2020). This approach combines methods and tools from text mining and social network analysis and relies on the analysis of word co-occurrence networks (Diesner, 2013) to calculate brand importance. The SBS indicator has three components – prevalence, diversity, and connectivity – more extensively described in the next section. While improving prevalence and diversity is usually easier for brand managers and practitioners – for example, requiring more social media posts that mention a brand name and create distinctive associations – the actions needed to improve brand connectivity are often not easy to guess. Connectivity is measured by the weighted betweenness centrality (Brandes, 2001) of a brand node in a semantic network. Its improvement requires linking a brand to specific nodes (words or concepts) that are strategically important to increase its "brokerage" power. This is more complicated than improving diversity, which only considers the number and uniqueness of brand associations. Specifically, we need to know which brand associations should be introduced or strengthened.

From a network science perspective, increasing connectivity means solving a MAXIMUM BETWEENNESS IMPROVEMENT (MBI) problem (D'Angelo et al., 2016) – which we extend to consider adversarial nodes, constrained budgets, and a weighted betweenness scenario where improvement is not only obtained by adding links but also considering the possibility of increasing the weight of existing connections. A more formal introduction to the MBI problem and the extension of interest for our research are provided in the following sections.

Our contribution is twofold. Firstly, we present an algorithmic framework to solve an extended version of the MBI problem that better fits the considered application scenario (i.e., adversarial and forbidden nodes, constrained budgets, and the strengthening of existing connections). An experimental analysis shows that our implementation is fast enough to apply to real-world scenarios effectively. Secondly, we integrate the framework into a new software system called *Brand Network Booster* (*BNB*), which supports scholars and practitioners who want to evaluate and improve brand connectivity. The system has an interactive user interface that allows users to easily customize the algorithmic framework based on their needs. Three case studies based on real-world data illustrate how the system can be used to improve brand connectivity in different scenarios.

The remainder of the paper is organized as follows. In the next section, we define and extend the MBI problem and link it to the measurement of brand connectivity. Subsequently, we present the main characteristics of our BNB software and the algorithmic framework behind it (Section 3). In Section 4, we present real-world applications. In the last section, we illustrate the main conclusions and implications for future research.

## 2. Solving an MBI problem to improve brand connectivity

### 2.1. Brand evaluation through semantic network analysis

Textual data analysis, semantic and socio-semantic networks have received increasing scholarly attention in recent decades (Chartier et al., 2020; Diesner, 2013; Park et al., 2024; Rust et al., 2021; Saint-Charles & Mongeau, 2018; Schöps & Jaufenthaler, 2024; Zhang & Moe, 2021). Many researchers presented novel methods and tools for analyzing these networks and showed the advantages of combining text mining with social network analysis (Fronzetti Colladon et al., 2020).

In particular, one of the most promising uses of text mining and network analysis has been in brand importance quantification. Based on these methods, Fronzetti Colladon (2018) introduced a metric known as the Semantic Brand Score to evaluate brand importance from textual data. According to this conceptualization, brand importance depends on the three dimensions of prevalence, diversity, and connectivity. The first is measured as a frequency count of the brand name in a text corpus, thus not requiring network analysis. On the other hand, diversity and connectivity depend on the consideration of textual associations and their patterns, being measured through network centrality indices – distinctiveness centrality (Fronzetti Colladon & Naldi, 2020) and weighted betweenness centrality (Brandes, 2001), respectively.

The Semantic Brand Score has proven helpful in predicting political election outcomes, forecasting stock and bond market returns and volatilities, quantifying climate risk exposure at the firm level, or measuring the relationship between brand value and revenues, among others (Fronzetti Colladon, 2020; Fronzetti Colladon et al., 2023; Rovelli et al., 2022; Vestrelli et al., 2024).

Despite the growing literature on text mining and network analysis, a notable research gap remains concerning the strategic enhancement of brand connectivity within semantic networks. While existing studies have focused on measuring dimensions such as prevalence and diversity, the issue of improving brand connectivity—defined as the strategic enhancement of brand associations and discourse pathways—has received comparatively less attention. This gap is particularly pertinent given the increasing importance of brand connectivity in shaping consumer/citizen perceptions and behaviors, as well as its potential impact on brand equity (e.g., Fronzetti Colladon et al., 2023; Rovelli et al., 2022).

From a practitioner's point of view, increasing diversity and prevalence is easier, as it requires that a brand's name is more frequently mentioned (for example, by opinion leaders or media) and associated with a richer discourse (with more, and less common, associations). By contrast, improving connectivity is a difficult task. It requires knowledge about all the possible paths in a network and identifying strategic associations to improve a brand's "brokerage power" in the discourse. Therefore, it is necessary to provide practitioners (and scholars) with specific suggestions on which words/concepts would be more relevant in improving brand connectivity – to guide the creation of new brand links or a weight increase of existing connections.

The system we built and present in the following sections has this exact scope, i.e., supporting the determination of the most strategic brand associations and simulating possible scenarios. To the extent of our knowledge, this is the first decision support system focused on improving brand connectivity, which also considers maximum budgets, possible adversarial nodes, and link weight change. Indeed, enriching brand associations has a communication cost that must be considered. Additionally, some companies might want to ensure that with their interventions, they do not modify network paths to favor competitors' brands. Similarly, they must ensure that associations align with the brand strategy, avoiding misaligned or detrimental connections. Regarding network theory, this corresponds to solving an extended version of an MBI problem (D'Angelo et al., 2016).

It is important to note that, even if the SBS conceptualization inspired our work, measuring a brand's betweenness centrality in semantic networks is interesting per se. Indeed, several other scholars looked at this measure and considered it a proxy of brand relevance and popularity (e.g., Gloor, 2007; Gloor et al., 2009). Other significant results can be found in the work of Netzer and colleagues (2012), who combined text mining with social network analysis to generate market structure perceptual maps and obtain strategic insights without the need to conduct consumer interviews. Also, these authors used betweenness centrality in their investigation. Lastly, our tool could support research concerning the calculation of semantic importance beyond brand studies.

### 2.2. MBI problem and related variants

Let $G = (V, E)$ be an undirected graph where $|V| = n$ and $|E| = m$. We assume $G$ to be *weighted*, i.e., $G$ comes with a function $\omega : E \to \mathbb{R}_0^+$. For each $e \in E$, the value $\omega(e)$ is the *weight* of $e$. Also, for each path $\pi$ of $G$, the *weight* of $\pi$ is the sum of the weights over all its edges. For any two nodes $s, t \in V$, a *shortest path* between $s$ and $t$ is a path having $s$ and $t$ as endpoints whose weight is the lowest among all such paths. The *distance* in $G$ between $s$ and $t$, denoted by $d_{st}$, measures the weight of any shortest path between them. Similarly, $\sigma_{st}$, and $\sigma_{st}(v)$, denote the number of shortest paths between $s$ and $t$ in $G$, and the number of shortest paths between $s$ and $t$ in $G$ that contain $v$, for some node $v \in V$, respectively. For each node $v$, $N_v$ denotes the set of neighbors of $v$, while the *betweenness centrality* of $v$ is defined as

$$b_v = \sum_{s,t \in V \setminus \{v\}} \frac{\sigma_{st}(v)}{\sigma_{st}}$$

For a set of edges $S \subseteq (V \times V) \setminus E$, we let $G_S = (V, E \cup S)$. We are inter-

ested in the following problem (D'Angelo et al., 2016):

**MAXIMUM BETWEENNESS IMPROVEMENT (MBI)**

**Input**: A weighted graph $G = (V, E)$, a node $v \in V$, an integer $k \in \mathbb{Z}^+$, a real $\delta \in \mathbb{R}^+$
**Problem**: find a set of edges $S = \{uv : u \in V \setminus N_v\}$, each having weight $\delta$, such that $|S| \leq k$ and the new value $b_v$ in $G_S$ is maximum.

**Algorithm 1 (MBI)**

**Input**: A weighted graph $G = (V, E)$, a node $v \in V$, an integer $k \in \mathbb{N}$, a real $\delta \in \mathbb{R}_0^+$
**Output**: A set of edges $S = \{uv : u \in V \setminus N_v\}$, each having weight $\delta$, such that $|S| \leq k$
1. $S \leftarrow \emptyset$;
2. **for** $i = 1, 2, \ldots, k$ **do**
3.     **foreach** $u \in V \setminus N_v | uv \notin S$ **do**
4.         compute $b_v(S \cup \{uv\}) : w(uv) = \delta$;
5.     $u_{max} = \mathbf{argmax}\{b_v(S \cup \{uv\}) : u \in V \setminus N_v \wedge uv \notin S\}$;
6.     $S = S \cup \{u_{max}v\}$;
7. **return** $S$;

Under standard complexity assumptions, the MBI problem cannot be approximated within a factor greater than $1 - \frac{1}{2e}$ (D'Angelo et al., 2016). On the other hand, D'Angelo et al. (2016) describe a greedy algorithm (see Algorithm 1) that often adds only a few edges to increase the betweenness centrality of a node, although its approximation ratio can be unbounded in some instances. In Section 3, we extend this algorithm to fit the application domain requirements anticipated in Section 2.1.

We remark that the MBI problem has also been studied for directed graphs (Bergamini et al., 2018), which is less useful for our purposes. Moreover, the parameterized complexity of the problem in the unweighted and undirected setting has also been investigated (Hoffmann et al., 2018).

## 3. System design and engineering

The design of the BNB software relies on the "Data-Users-Tasks"' model proposed by Miksch and Aigner (2014). Following this model, we describe the data, the users, and the tasks behind our design process (Section 3.1). Subsequently, we describe the algorithmic framework at the system's core (Section 3.2) and the visual interface design (Section 3.3). Details about the implementation and its performance can be found in the appendix.

### 3.1. User-Data-Tasks

The *users* of our system are scholars and practitioners who are experts in their domain but are not expected to have a background in algorithm design and programming. For example, they could be social network scholars, managers, and analysts working in a company's marketing or communication department.

In terms of *data*, the input of our system is a word co-occurrence network generated by the SBS BI software (Grippa & Fronzetti Colladon, 2020) or available in the popular Pajek NET format (De Nooy et al., 2018). In what follows, we will refer to a word co-occurrence network given in input to the system simply as the input network. In particular, the input network is a weighted undirected graph in which edge weights are always integers, representing the number of co-occurrences of each word pair in a corpus.

The system has been designed with the following two main analysis *tasks* in mind.

*T1: Improve the connectivity of a brand.* The user specifies a *target node* (a brand) in the input network, and the goal of the task is to increase its betweenness centrality by inserting new edges in the network with suitable weights or by modifying the weights of existing edges. The only structural constraint is that inserted or modified edges be incidents to the target node.[2] Since enriching brand associations has a communication cost, the user should be able to provide a budget expressed in terms of the maximum total increment considering all edge weights. In addition, some companies might want to ensure that their interventions do not modify network paths in a way that favors competitors' brands. Accordingly, the user should be allowed to specify a list of nodes corresponding to such competitors. For example, while increasing the connectivity of "Coca-Cola," practitioners would probably desire not to favor "Pepsi." Moreover, companies must ensure that associations are not created that are not aligned with the brand communication strategy. Consequently, the user should be allowed to specify a list of nodes for which a connection with the target node is forbidden. For example, a company selling toys for kids would probably like to have the word "war" outside its brand image.

---

[2] In the following we will use the term node and vertex interchangeably.

*T2: Experiment with the effect of adding new links to the brand.*
The user should be allowed to specify a pair of nodes and a weight in the input network to simulate the effect of new potential associations. This task aims to experiment with the result of adding an edge between two nodes with a given weight. In particular, the user should be able to evaluate the effect on the betweenness centrality of the two selected nodes.

### 3.2. Algorithmic framework

Based on task *T1*, we formalize a new variant of the MBI problem that better fits the needs of our application. We call this variant CO-MBI (CO-OCCURRENCE MAXIMUM BETWEENNESS IMPROVEMENT). Let $G = (V, E)$ be the weighted undirected input graph, with $V$ being the set of vertices and $E$ of edges. To simplify the problem statement below, we let $K(G)$ be the complete weighted graph over vertex set $V$ such that the weight of an edge $e$ is equal to the original value $\omega(e)$ if $e \in E$ and it is 0 otherwise (i.e., if an edge does not exist in $G$). Consider a set of edges $S \subseteq (V \times V)$ and a function $\omega_S : S \to \mathbb{Z}^+$; we let $K(G, S, \omega_S)$ be the graph obtained from $K(G)$ by increasing the weight of each edge $e \in S$ by $\omega_S(e)$. Also, we call the *cost* of $\omega_S$ the value $\sum_{e \in S} \omega_S(e)$.

**CO-OCCURRENCE MAXIMUM BETWEENNESS IMPROVEMENT (CO-MBI)**

**Input**: A weighted graph $G = (V, E)$, a target node $v \in V$, an integer budget $k \in \mathbb{Z}^+$, a subset of opponent nodes $C \subset V$, a subset of forbidden nodes $F \subset V$.
**Problem**: find a set of edges $S = \{uv : u \in V \setminus F\}$ in $K(G)$ and a function $\omega_S : S \to \mathbb{Z}^+$ of cost at most $k$, such that (1) the new value $b_v$ in $K(G, S, \omega_S)$ is maximum, and (2) the new value $b_c$ in $K(G, S, \omega_S)$ for any $c \in C$ is not larger than in $K(G)$.

In order to solve the CO-MBI problem, we suitably modify and extend the pseudocode shown in Algorithm 1, as reported in Algorithm 2. The main differences are the exclusion of forbidden nodes (line 8), a binary search (line 10) to efficiently find the minimum weight that leads to a significant increase of betweenness centrality, the additional conditions imposed by opponent nodes, i.e., brand competitors, (lines 16–18) and budget constraints (line 3). The variable $p_{imp}$ is initialized to 0.01 but can be adjusted by the user based on the adopted strategy and the desired betweenness gain of the opponents that the user is willing to accept.

**Algorithm 2**

**Input**: An undirected graph $G = (V, E)$; a target node $v \in V$; a budget $k \in \mathbb{Z}^+$; a set of forbidden nodes $F$ and a set $C$ of opponent nodes.
**Output**: A set of edges $S = \{uv : u \in V \setminus F\}$ in $K(G)$ and a function $\omega_S : S \to \mathbb{Z}^+$
1. $S \leftarrow \emptyset$;
2. $\omega_s(e) = 0 \forall e \in E$;
3. **while** $cost(\omega_s) \leq k$ **do**
4.     $b_{v-max} \leftarrow b_v in K(G, S, \omega_s)$;
5.     $b_{v-best} \leftarrow b_v in K(G, S, \omega_s)$;
6.     $u_{best} \leftarrow \emptyset$;
7.     $\omega_{best} \leftarrow 0$;
8.     $p_{imp} \leftarrow 0.01$;
9.     **foreach** $u | u \neq v, u \neq s \forall sv \in S, u \notin F$ **do**
10.         $\omega_t \leftarrow \omega_s$;
11.         $B \leftarrow binarySearch(k - cost(\omega_s), b_v in K(G, S, \omega_t))$;
12.         $b'_v \leftarrow b_v in K(G, S, \omega_t)$;
13.         $\omega'(uv) \leftarrow \omega(uv) in K(G)$;

(*continued*)

| Algorithm 2 | |
|---|---|
| 14. | **while** $B.continue(b'_v)$ **do** |
| 15. | $\omega_t(uv) \leftarrow B.nextWeight()$; |
| 16. | Compute $b''_v inK(G, S \cup \{(uv)\}, \omega_t)$; |
| 17. | **foreach** $c \in C$ **do** |
| 18. | Compute $b_c inK(G, S \cup \{(uv)\}, \omega_t)$; |
| 19. | **if** $(b''_v - b'_v) \geq (p_{imp} \times b'_v)$ **and** $\forall b_c$ respects the condition |
| 20. | $b'_v \leftarrow b''_v$; |
| 21. | $\omega'(uv) \leftarrow \omega_t(uv)$; |
| 22. | **if** $b'_v \geq b'_{v-best}$ |
| 23. | **if** $b'_v > b_{v-max}$ |
| 24. | $b_{v-max} \leftarrow b'_v$; |
| 25. | **if** $\left((b'_v - b_{v-best}) \geq (p_{imp} \times b'_{v-best})\right)$ **or** $(\omega'(uv) \leq \omega_{best})$ |
| 26. | $b_{v-best} \leftarrow b'_v$; |
| 27. | $u_{best} \leftarrow u$; |
| 28. | $\omega_{best} \leftarrow \omega'(uv)$; |
| 29. | **else** |
| 30. | **if** $(b_{v-max} - b'_v) < (p_{imp} \times b'_v)$ **and** $(\omega'(uv) < \omega_{best})$ |
| 31. | $b_{v-best} \leftarrow b'_v$; |
| 32. | $u_{best} \leftarrow u$; |
| 33. | $\omega_{best} \leftarrow \omega'(uv)$; |
| 34. | **if** $u_{best} == \emptyset$ |
| 35. | break; |
| 36. | $\omega_s(u_{best}v) \leftarrow \omega_{best}$; |
| 37. | $S \leftarrow S \cup \{(u_{best}v)\}$; |
| 38. | **return** $S, \omega_s$; |

### 3.3. Data visualization and user interaction

We developed a visual interface based on the following views to support an effective interaction between the users and the algorithmic framework.

#### 3.3.1. Overview

Fig. 1 shows the system homepage. The user can load a network, and the system will show an overview of such a network.[3] The system exploits a technique to render massive node-link layouts with adjustable levels of abstraction interactively. This is especially useful to get an initial view of the network structure. As the underlying layout algorithm is based on a force-directed model, words (and hence brands) that are well connected tend to be placed in a central position in the visualization. In contrast, loosely connected words occupy more peripheral positions. Zooming in on specific portions of the visualization allows the user to explore the corresponding subgraph in more detail, looking, for example, at explicit connections between pairs of words.

#### 3.3.2. Connectivity Improvement view

This view is designed to support task T1. The user can set all the parameters of the algorithmic framework developed to solve the CO-MBI problem. Namely, the user can select a target node (brand), a budget, a set of opponent nodes, and a set of forbidden nodes. Using a constrained budget is necessary to model the fact that increasing the association between two words (i.e., strengthening the brand image) might have a communication cost – for example, implying additional activities of social media influencers or the communication department of a company.

In addition to these parameters, there are other options we have added to the algorithm from which the user can choose. First, the user can determine whether the algorithm should be allowed to add new edges to the graph or should be limited to only increasing the weight of the existing edges. Second, the user can ask the system to avoid creating new edges toward nodes with low degrees. Third, the user can choose between multiple strategies concerning opponent nodes. The default option (NO-INCREMENT) implies that the betweenness centralities of opponents do not increase. Another strategy (UPPER-BOUND) allows us to define an upper bound to the improvement of the betweenness centralities of opponents. A third strategy (DELTA) tolerates only a maximum percentage increase (e.g., 5 %). The last strategy (DELTA-RATIO) ensures a minimum ratio between the growth of the betweenness centrality of the target node and the maximum growth of any opponent.

Once the algorithm has computed a solution, it shows the following information. For each added or modified edge, the user can see how the corresponding weight has varied and how much of the budget has been invested to achieve this result. Regarding opponents, it is shown how much their betweenness has varied as a percentage increase (or decrease).

#### 3.3.3. Shortest Paths and Connectivity Test views

In the Shortest Paths view, the user specifies a source and a destination node, and the system computes all the shortest paths between them. If the CO-MBI algorithm has already been executed, the system shows the shortest paths before and after adding new edges or changing the weights of existing edges.

The Connectivity Test view is tailored to support task T2. The user can select two nodes, and the system will display the current weight of the edge connecting them, which will be zero if there is no direct connection. The user can then modify this weight (either increase or decrease it) and observe the resulting changes in the betweenness centrality of the two nodes.

## 4. Case studies

In this section, we present three different case studies to demonstrate the potential and robustness of the software. Specifically, these case studies were selected and developed based on two criteria. First, to showcase the software's flexibility in handling various textual sources and brands. To do so, we analyzed semantic networks created from different corpora and focused on multiple and heterogeneous brands/ terms. Secondly, in each case study, we carefully selected situations that would validate the accuracy of the BNB output. These analyses were chosen to examine scenarios that occurred after the connectivity (betweenness) improvement.

Before constructing each semantic network, we cleaned the text by eliminating stop words, punctuation, and special characters. Furthermore, we converted all text to lowercase and extracted stems by removing word affixes (Porter, 2006). These procedures were executed using SBS BI software (Grippa & Fronzetti Colladon, 2020). We combined single words into bigrams, or trigrams, where appropriate, utilizing the Spacy library (Schopf et al., 2024). This approach ensures that instances such as "European Central Bank," "leverage ratio," or "cash flow" will be represented as a single node rather than three separate nodes. Accordingly, each node represents a word (or word combinations), and links exist if two terms appear close to one another in the text. Links are weighted based on their frequency of co-occurrence, with negligible co-occurrences (low frequency) being disregarded as they are likely due to chance rather than indicating a significant relationship among the words.

### 4.1. European central Bank communication

This case study focuses on the European Central Bank (ECB) response to fighting COVID-19. The analysis aims to show how the output of BNB would have aligned with policymaker decisions in response to the COVID-19 spread, suggesting the potential benefits of utilizing BNB software in identifying and guiding central bank communication during a pivotal moment in the early stages of the global pandemic. The target node of each analysis performed with BNB is "European Central Bank."

---

[3] The network visualization is computed by using a customization of a recent JavaScript-based system, called BrowVis (Consalvi et al., 2022).

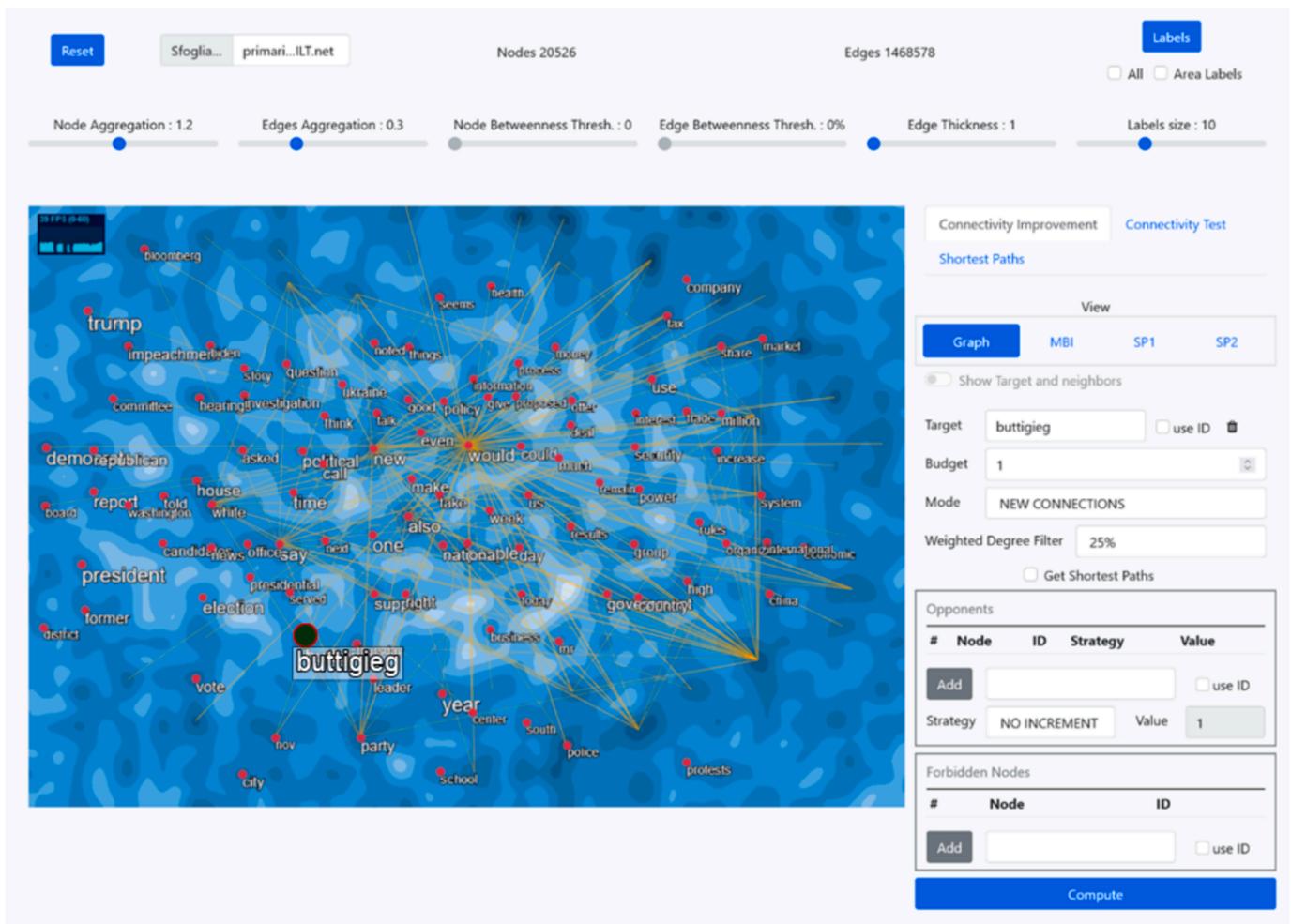

**Fig. 1.** Screenshot of the system homepage.

Each simulation is run to improve its betweenness centrality in a semantic network generated from news articles from March 12 to March 17, 2020.

*4.1.1. Background*

The European Central Bank (ECB) is a critical institution within the European Union (E.U.) and the broader global financial system. Established in 1998, the ECB's primary mandate is maintaining price stability within the Eurozone. The European Central Bank was among the first E. U. institutions to react to the pandemic outbreak, playing a crucial role in shaping the E.U.'s initial economic response to the crisis. One of the earliest interventions occurred in March 2020, adopting unconventional monetary policy tools to stabilize financial markets (Forman & Mossialos, 2021; Quaglia & Verdun, 2023). Under the leadership of President Lagarde, during the policy meeting of the Governing Council held in Frankfurt, the ECB promptly implemented an expansionary monetary policy toolkit aimed at contrasting the negative shock caused by the spread of the virus (ECB, 2020).

However, the market did not deem this policy intervention as sufficient to fight the pandemic for two main reasons: firstly, for the perceived inadequacy of the intervention in monetary terms, and secondly and most importantly, for the communication errors made by Lagarde during the press conference held on the same day (Financial Times, 2020; Reuters Staff, 2020). Indeed, during the press conference, responding to a question by a journalist, Lagarde stated that it was not the ECB's role to close spreads, emphasizing that other tools and actors were available to address such issues, contradicting the message she had given in the formal presentation prior (Quaglia & Verdun, 2023). These statements immediately led to financial turbulence, a deterioration in sustainability expectations for countries, and significant capital flight, resulting in increased spreads and a drop in stock market indices (Ganie et al., 2022). Lagarde's comments were interpreted as lacking support for the euro area countries, particularly for countries in the south, such as Italy (Quaglia & Verdun, 2023).

Lagarde gave another interview soon after and apologized for the mistake (Arnold, 2020; Clinch, 2020). Only six days later, on March 18, 2020, Lagarde called for an extraordinary Governing Council meeting to address the urgent situation (Jones, 2020). During this meeting, the ECB introduced the Pandemic Emergency Purchase Program (PEPP), a novel temporary asset purchase program for private and public sectors (ECB, 2020; Verdun, 2022).

The decisions and actions embarked on by the ECB during the second intervention exhibited strong support for E.U. member states, showcasing a coordinated approach between monetary and fiscal policies. Despite the ECB's stated independence, ECB actions and decisions on March 18 symbolized its willingness to counter the pandemic and support member states with the right mix of monetary policies. This intention was further supported on April 9, when the ECB's President Lagarde strongly called for "full alignment of fiscal and monetary policies" (European Paliament, 2020).

*4.1.2. Research question*

The ECB intervention on March 18 played a crucial role in stabilizing financial markets. What insight the BNB software would have provided if used during the week preceding March 18 by the ECB? Could it have provided valuable guidance to policymakers on the most effective intervention strategies? The research question explored in this case study aims to determine whether the ECB could have made sound policy decisions following the press conference and the communication error that occurred on March 12, supported by the BNB. To answer this research question, we use the BNB software combined with the information in online news articles the week before the press conference on March 18. This time frame and scenario are particularly suitable for analysis for several reasons. In particular, effective and credible communication is essential to a policy's success. Communication is an integral tool (if not the most important) of any monetary policy, and each ECB's communication passes through the news media before reaching the financial markets (Picault et al., 2022); therefore, analyzing news coverage during this period is valuable, especially considering the uncertainty and instability experienced after March 12.

Secondly, and most importantly, we have a scenario in place to test the empirical validity of the BNB output. The policy decisions made on March 18 were successful in stabilizing financial markets. After the meeting, there was a significant drop in term spreads of government bonds for the European countries most affected during the previous week, especially Italy. As a result, we expect the BNB output to align with the policy decisions made on March 18, suggesting expansionary actions and support for fiscal policy. Those actions effectively reduced spreads and addressed the government bond crisis more in general.

*4.1.3. Data*

For this analysis, we utilized news articles from U.S. news journals to capture the general discourse surrounding the European Central Bank's (ECB) monetary policy decisions. We sourced articles from CNN, The Financial Times, Bloomberg, Los Angeles Times, New York Times, and The Washington Post to ensure diverse and complete perspectives about ECB monetary policy decisions reported in news media articles. We gathered news articles published between March 12 and March 17, 2020, from the EventRegistry database (Leban et al., 2014). We retained only the articles that included the keywords "ECB" or "European Central Bank" in the title or body. We collected a total of 162 articles that met our specified criteria. The semantic network comprises 1115 nodes and 9328 links.

*4.1.4. Results and discussion*

As explained in the introduction, the target node is "European Central Bank." We produce several analyses to demonstrate the sensitivity of the results produced by the software using different budgets. We identified the most frequent nodes (words) produced by the BNB output through multiple simulations, starting with a budget of 100 and gradually increasing by 100 units until 1300 for thirteen different simulations. The strategy adopted in each simulation was "Change Weight." We opted for this strategy mainly because after the March 12 conference, the ECB already opted for an expansionary monetary policy. What remained unclear was the likelihood of supporting governments and having a more "accommodating" monetary policy. In other words, the target node of the analysis was already very central in the discourse and had numerous connections with various significant nodes. We also tried adopting the ''New Connection'' strategy as a robustness check. However, this led to only a few significant nodes throughout the analyses, namely ''oil,'' ''recession, '' and ''Trump.'' These nodes could indicate general macroeconomic risks related to the spread of COVID-19 and its impact on the energy sector. However, since we were interested in analyzing the monetary policy actions the ECB would need to undertake, we opted for the ''Change Weight'' strategy to analyze where the ECB should have placed more emphasis.

After each analysis, we considered the suggested key nodes and constructed Table 1 according to the number of times each node appeared among the results. Additionally, we report in Fig. 2 a word cloud of the most common terms observed in the output of the runs.

**Table 1**
BNB output.

| Word | Count | Category |
|---|---|---|
| Cash | 12 | *Policy Actions* |
| Fiscal | 11 | |
| Policy | 3 | |
| Job | 9 | *Pandemic Effect* |
| Global | 4 | |
| Travel | 3 | |
| Lagarde | 6 | *Communicatoin Error* |
| Debt | 2 | |
| Italian | 2 | |
| President | 1 | *Residuals* |
| Bank | 1 | |
| Swiss | 1 | |

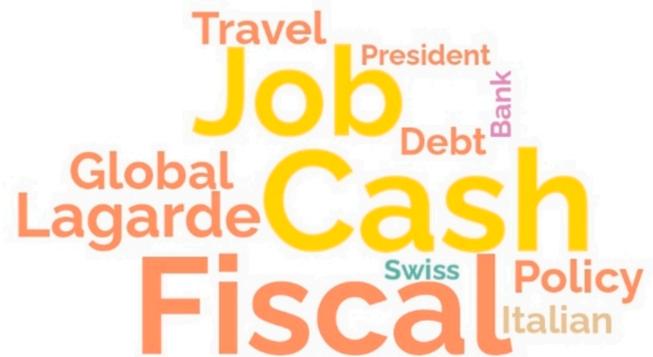

**Fig. 2.** Word cloud of most important targeted nodes.

Bigger words are those more frequently observed in the output. From this sensitivity test, we note that the BNB output heterogeneity diminishes while increasing the budget until reaching a "stagnation" point after surpassing the threshold of 1000.

From a practical point of view, two different strategies could be adopted. The first is to run numerous simulations and discard nodes that appear below a certain threshold (for example, targeted nodes that occur only in one or two simulations, in our case, "Swiss," "bank," and "president"). In contrast, the second strategy is to increase the budget until a "stagnation" point is reached. During the analyses, after exceeding the budget of 1000, the key nodes remained the same.

After performing these simulations, we rationalized the targeted nodes into three (plus a residual) different categories. The first category, "policy actions," is made up of the pivotal nodes "cash," "fiscal," and "policy." The first two nodes of this category are the most robust, considering all runs. They appear 12 and 11 times, respectively, confirming their importance. The nodes "cash" and "fiscal" may suggest that emphasizing liquidity provisions and coordination with fiscal policies could have reassured markets of the ECB's commitment to stabilizing the economy.

Not by chance, on March 18, the ECB announced a new temporary program for public and private sector asset purchases with an envelope of EUR 750 billion and with a high degree of flexibility (European Parliament, 2020). As a matter of comparability, on March 12, the Governing Council decided to increase the existing Asset Purchase Program (APP) net purchases by "only" EUR 120 billion.

Additionally, on March 18, the ECB decided to include assets from non-financial corporations in the Corporate Sector Purchase Programme (CSPP), showing great support and alignment with fiscal policies. Overall, the monetary policy decisions made one week later seem to align with the suggested nodes "cash" and "fiscal" provided by the BNB output.

The second group of terms consists of "job," "global," and "travel,"

which are likely to reflect the impact of COVID-19 on the economy. Both "job" and "travel" appear to reflect public concerns regarding employment and travel restrictions resulting from the pandemic. Similarly, the term "global" suggests the widespread effects of the coronavirus. By examining the most common associations that occur in the news article texts with "global," we find terms such as "outbreak" and "coronavirus".

This second group of keywords does not provide a clear direction for ECB policy actions. Still, it acknowledges the importance and relevance of the impact that COVID-19 was having on the economy at that time. It may suggest that policymakers should have directly intervened to limit the virus's spread. While the first group of keywords seems to describe "how" policymakers should have acted, the second group seems to indicate "what" they should have counterbalanced.

The last group, made up of "Lagarde," "debt," and "Italian," seems to relate to the communication errors made by the president during her previous speech. Not by chance, one of the countries most affected by the financial turbulence of those days was indeed Italy, creating serious concern for public debt sustainability. Together with these two keywords (debt and Italian), the node "Lagarde" is consistent across six different simulations. Once again, the BNB output resembles what happens in subsequent days. On March 18, 2020, Lagarde posted a tweet saying: "Extraordinary times, require extraordinary measures, […] our commitment to the euro is unlimited." Those words looked like a reiteration of Mario Draghi's statement that the euro was irreversible and the ECB would do whatever it takes to that effect. The bond spread of the most affected E.U. countries, especially Italy, started dropping after those declarations and the policy intervention on March 18. We classify the key nodes "President," "Swiss," and "bank" as residual because they appear only once during all the simulations.

Overall, the results obtained from the simulations are quite in line with the ECB's March 18 actions. The BNB seems to suggest three key macroeconomic issues. The first concerns the monetary policy actions the ECB should have implemented to address a potential new spread crisis. The second relates to the actions the ECB should have taken to address the global pandemic. Lastly, an urgent need for an intervention to rectify communication errors. Actions aligned with these findings were largely implemented on March 18, resulting in positive outcomes such as reduced spreads and increased stability in financial markets.

*4.2. U.S. Democratic Party presidential primaries*

This case focuses on the 2020 Democratic Party presidential primaries in the United States, focusing the analysis on one specific candidate: Joe Biden. The study aims to validate and show how BNB results would have aligned with the leverage points used by Biden during his electoral campaign, suggesting some potential benefits of using BNB software in bolstering political brands. The target node for each analysis performed with BNB is "Joe Biden." Each simulation is run to improve the betweenness centrality of "Joe Biden" in a semantic network generated from news articles from February 25, 2020, to March 2, 2020.

*4.2.1. Background*
In the United States, before the presidential election (which in 2020 saw former Vice President Joe Biden competing against Donald Trump), candidates for each party—Democratic and Republican—are selected. The timeline for the 2020 Democratic Party presidential primaries spanned from February 3 to August 11, 2020. During that period, voters in each state cast their ballots for one of the candidates, either through primaries or caucuses. However, not all states hold their votes on the same day.

Our focus will be on one of the most significant events of the primaries: Super Tuesday. Super Tuesday is a critical event in the primary season (Phillips, 2020). It is so-called because it is the day when the most significant number of states hold their primaries or caucuses. The outcomes of Super Tuesday often align with the primaries' final results (Phillips, 2020).

We decided to focus our analysis on the days before the 2020 Super Tuesday because of the unique situation that had arisen. After February 25 (one week earlier than Super Tuesday, which was on March 3), the candidates running for Democratic primaries were Bernie Sanders, Joe Biden, Pete Buttigieg, Amy Klobuchar, Michael Bloomberg, Elizabeth Warren, and Tulsi Gabbard. Before Super Tuesday, Bernie Sanders was the frontrunner, winning key events and dominating expectations. On February 3 in Iowa, Sanders secured a plurality of votes in both the first and final alignments. On February 11, he won New Hampshire with 26 % of the vote, with Buttigieg trailing closely at 24 %. Sanders continued his winning streak on February 22 in Nevada. Biden's first victory came only on February 29 in South Carolina, three days earlier than Super Tuesday.

Biden's triumph on Super Tuesday marked one of the greatest comebacks in political history (Collinson, 2020; Stevens & Grullón Paz, 2020). Despite Sanders being the favored candidate (Enten, 2020), both in perception and numbers, the results on March 3 were a game-changer. Biden won in Alabama, Arkansas, Massachusetts, Maine, Minnesota, North Carolina, Oklahoma, Tennessee, Texas, and Virginia, while Sanders secured victories in California, Colorado, Utah, and Vermont. What fueled this remarkable comeback? More importantly, would the BNB model's output, based on the news in the week following Super Tuesday, have provided actionable insights for Biden's resurgence?

*4.2.2. Research question*
In retrospect, it is possible to argue that the key event that catalyzed Biden's comeback was his victory in South Carolina. The victory anointed Biden as a frontrunner by helping him gain support among African American voters, furthering his success in Southern states (Collinson, 2020; Stevens & Grullón Paz, 2020). What guidance would the BNB software have provided if it had been used by Joe Biden in the week leading up to Super Tuesday? Would it have suggested stronger ties to South Carolina or other states? The research question explored in this case study seeks to determine if the BNB's analysis would have advised Biden to leverage his victory in South Carolina. Given the substantial impact of this victory on the election results, generating an output specifically for South Carolina, rather than linked to other factors, would provide some evidence about software potential.

*4.2.3. Data*
As Fronzetti Colladon (2020) highlights, the most informative period for predicting election outcomes is the week before the voting event. Consequently, we collected all the news articles related to the primaries from the Event Registry database for the week preceding Super Tuesday, specifically from February 25, 2020, to March 2, 2020. The news articles were filtered using specific keywords, including "presidential primary," "democratic primary," "super Tuesday," and the names of the candidates to ensure comprehensive coverage of the election narrative. As in the previous case study, the newspapers used for this analysis include CNN, The Washington Post, Financial Times, The New York Times, Los Angeles Times, and Bloomberg, ensuring spatial homogeneity by including publications from both the West and East Coasts. After filtering the news, we got 508 news to create the semantic network.

In this second case study, to demonstrate the flexibility of the BNB model, we generated the semantic network using an entity linking (EL) system which links entity mentions in documents to their corresponding entities in Wikidata (over 30 M entities) (Ayoola et al., 2022). Unlike a semantic network formed solely by words, the EL system enables us to link and disambiguate each entity within the text. For example, if the text contains the word 'Joe' referring to Biden, the EL system will link this word to the candidate. Generally, using EL reduces the number of nodes in the network by merging differently written entities that refer to the same concept.

We used the entity linking system for two main reasons. The first is accuracy. Given the nature of the case study and the high number of candidates, greater precision and disambiguation of each candidate's mentions make the formation of the semantic network more accurate. Additionally, we want to highlight that it is possible to build and operate on semantic networks created in different ways, even using more complex methods when necessary. The network comprises 1575 nodes and 20,354 links.

*4.2.4. Results and discussions*

As previously mentioned, the target node of this study is Joe Biden. As with the previous case study, we started with an initial budget of 100. We increased it by 100 units until reaching a point of stagnation. This resulted in a budget of 1500. We conducted 15 different simulations (we tried simulations with much higher budgets, such as 5000, but the results were not substantially different from 1500).

We investigated two different scenarios. In the first phase, we ran simulations without including opponent nodes (Table 2, column 2). In the second phase, we included the remaining candidates as opponent nodes[4] (Table 2, column 3). We opted for the "Both" strategy in this case study. This allowed the model to recommend either reinforcing existing connections to the target node or establishing connections to new nodes not currently linked to the target node.

When the budget is low, specifically below 600 units, the most frequent node is "South Carolina." Once the budget exceeds 600 units, "South Carolina" disappears, and the most frequent terms become "election" and "state." Some residual nodes appear jointly during simulations. These include "politic" (3 times), "letter" (2 times), "Sanders" (2 times), and "Buttigieg" (1 time). However, since they do not robustly appear during many simulations, we exclude these nodes from the discussion.

The analysis highlights that with a limited budget, the BNB model directs attention to strategically significant yet accessible connections like South Carolina. Conversely, with a larger budget, the model focuses on broader electoral strategies, emphasizing the pursuit of support from various states and a wider voter base.

From a media and communication perspective, it would have been (and indeed was) very easy and advantageous for Biden to strengthen the connection with South Carolina, given his recent victory there. Logically, Biden would find it much easier to link to South Carolina rather than to other states or voters in general, as suggested by higher-budget simulations. The BNB results with higher budgets are reasonably obvious. During elections, it is natural for a candidate to seek support from other states and new voters. Although this is straightforward, it shows that the BNB output reflects the analyzed scenario.

After introducing opponent nodes and opting for the "No Increment" strategy, the system failed to find any solution when the budget surpassed 200 units. Consequently, we allowed the system more flexibility by opting for a "Delta" strategy. As reported in Table 2, results using opponents are identical to those obtained without opponent nodes. Yet, analyzing how the betweenness centrality of opposing candidates varies is interesting.

Indeed, one of the software's advantages and contributions lies in its ability to analyze the variation in connectivity of competitors during each simulation when they are included in the set of opponent nodes.

During the simulations, all candidates experience a reduction in connectivity. The only exception occurs for Amy Klobuchar, who undergoes an almost negligible average increase in connectivity, amounting to 0.15 % in simulations where the output node is ''South Carolina''.

This case study suggests that the BNB software would have advised the candidate to strengthen their connection with the South Carolina node to take on a more central role in the semantic network. This advice was later validated in reality. In hindsight, Biden's victory in South Carolina was pivotal in his campaign. Another important aspect to consider is the budget's impact on the output. Results differ depending on whether the budget is high (above 600 units) or low (below 600 units). While the recommended nodes vary, the underlying concepts are similar. Both the "South Carolina", the "election" and "state" nodes probably suggest the need for stronger connections with a wider electoral base.

The key difference is that in low-budget simulations, the suggested node is closer to the candidate, both geographically and in media coverage, following his recent victory in that state. In this case study, with a lower budget, the software advises focusing on words that are more immediately accessible to the candidate. In other words, it is easier for Biden to strengthen his connection to the name of a state where he recently won and has high media visibility rather than strengthening ties with other states.

In contrast, with a higher budget, the software suggests stronger connections with nodes that are more general and challenging to reach ("state" and "election") but potentially more impactful in the long run. This would be a broader strategy, connecting to terms aiming at a more expansive and far-reaching influence.

*4.3. Covid-19 firm exposure*

This case study aims to showcase the effectiveness of BNB software as a monitoring tool for policymakers in the initial phases of the COVID-19 pandemic. To demonstrate the adaptability of the BNB to deal with semantic networks from different textual sources, we decided to conduct this case study using company earnings conference call transcripts rather than online news articles. The target node in this case study is "COVID-19," and the semantic network was built using the transcripts of conference calls held in March 2020 by all the U.S. firms in the Refinitiv Eikon database (https://eikon.refinitiv.com/) available for that period.

*4.3.1. Background*

One of the most significant challenges regulators and policymakers face when dealing with unexpected and large shocks, such as the COVID-19 pandemic, is the urgent need to assess the impact on businesses. However, several measurement challenges must be overcome to address shock impacts on firms. Policymakers must determine how to identify affected firms, quantify the extent of firms' exposure to the shock, and understand the nature of the impact (e.g., demand contraction, supply chain disruption, financial difficulties) (Hassan et al., 2023).

A promising approach to quantify firms' exposure to such shocks, pioneered by Hassan et al. (2019), lies in earnings call transcripts analysis. Earnings calls are crucial events in the investor relations calendar where executives provide insights into the company's performance and respond to questions from financial analysts and other stakeholders. During those discussions, investors receive valuable insights into the company's financial health and current market trends (Hollander et al., 2010). Earnings calls are held quarterly by publicly traded companies and can be accessed through the Refinitiv Eikon database.

To assess a firm's exposure to a shock (e.g., COVID-19), Hassan et al. (2023) suggest measuring how central and critical the discourse about the shock (COVID-19) is during the call meeting discussions. The logic behind this measurement approach is simple: the more a company talks and centers the debate toward a specific topic, the higher the probability it will be affected by that event. An entire branch of literature showed that the more attention a topic receives during calls, the greater its potential impact on the company. Recent studies have validated this methodology for assessing a company's exposure to various risks, including political risk, Brexit, climate change, and significant events like the Fukushima nuclear disaster (Hassan et al., 2019, 2023; Sautner et al., 2023).

**Table 2**
BNB Output according to different budgets.

| Target | No Opponents | With Opponents |
| --- | --- | --- |
| South Carolina | 100–600 | 100–600 |
| State | 700–1500 | 700–1500 |
| Election | 1000–1500 | 1000–1500 |

---
[4] Excluding Tulsi Gabbard, who did not appear in the semantic network.

*4.3.2. Research questions*

The purpose of this case study is not to measure the extent of exposure to COVID-19 among the companies in our sample. This exercise could be achieved through a bunch of methods, such as frequency counts of specific terms (e.g., "COVID-19″ or "pandemic") or advanced text mining algorithms (such as word embedding models) on call transcripts.

On the other hand, our focus here is to use the BNB to identify the key links that, if established or strengthened during the initial phase of the pandemic, would make the COVID-19 term more central in the discourse. This means using the BNB to determine which associations or relationships involving COVID-19 would be crucial in making it a dominant topic in conversations. In hindsight, we know that in the following months, the COVID-19 term became extremely central in company discourses for various reasons, including the oil price shock (Hassan et al., 2023). Consequently, if the BNB output included nodes related to oil prices, it would contribute to proving the usefulness of the software.

The usefulness of this exercise can be seen in two different aspects. First, conducting an analysis with the BNB on conference calls at the time "t" can identify a group of key connections that can potentially enhance COVID-19's connectivity. A higher level of this term connectivity implies a greater level of risk. Therefore, at the time "t," policymakers would get information about which specific connections to monitor to mitigate or reduce the risks that firms may face in subsequent periods. This analysis would enable policymakers to determine which areas to focus on over different periods. Identifying nodes that can elevate COVID-19's centrality beforehand is crucial, as a more central role indicates greater company exposure to this risk.

Second, conference call transcripts are at the company level. This is an extremely important factor as it allows us to quantify how much a single company contributes to strengthening the connectivity of the "COVID-19″ term in the semantic network. For example, if the BNB output includes "crude oil," one can identify the company reports where that term appears most frequently. Once this is done, monitoring a specific company to see what happens if the COVID-19 term rises in connectivity due to a stronger connection to the term "crude oil" during subsequent periods is possible.

*4.3.3. Data*

As mentioned above, we collected conference call transcripts from the Refinitiv Eikon database. We gathered data from all the U.S. companies available for the March period, resulting in 763 overall transcripts from March 1 to March 31. The semantic network comprises 2069 nodes and 27,106 links. This timeframe was selected for two primary reasons.

As presented in the study by Hassan et al. (2023), discussions related to COVID-19 began to occur in early 2020; however, it was only after March that they experienced a surge. Performing the analysis in the months before March would have resulted in a too low frequency of "COVID-19." Similarly, repeating the analysis in subsequent months would not have allowed us to verify our results ex-post. In other words, after April, discussions were very much focused on COVID-19; our goal here is to demonstrate how the policymaker could have performed this analysis in a scenario that occurred at the early stage of the information propagation about the pandemic.

*4.3.4. Results and discussion*

In the first phase, we conducted simulations using a "Change Weight" strategy. We identified the most common nodes through multiple simulations, starting with a budget of 100 and then gradually increasing it until a stagnation point, consisting of a budget of 1000, totaling ten different simulations.

Out of all the simulations conducted, only two nodes consistently appeared in over 80 percent of cases: "concern" and "spread." Other nodes did occur consistently throughout the simulations. These additional nodes often included common terms like "thought" or "staff" or terms such as "crisis," "portfolio," "China," "clinical trial," and "communities." However, due to their high variability, we focus the discussion exclusively on the most recurrent terms, as done previously.

The words "concern" and "spread" clearly reflect the early stages of the pandemic, when companies were grappling with uncertainty about the virus's progression, and a general sense of concern prevailed.

In the second phase of the analysis, we carried out a series of simulations on the semantic network using a different strategy. Instead of employing the "Change Weight," we opted for the "New Connections" strategy, which allowed the creation of new links with COVID-19. We conducted 30 distinct simulations overall, starting with a budget of 100 and progressively increasing it to 1000, where we observed a stagnation point. For each budget, we executed three distinct analyses using weighted degree filters of 25 %, 50 %, and 75 % to test for the robustness of the output, i.e., we excluded low-weighted-degree nodes from the set of potential target nodes. We used various weighted degree filters because we are employing the ''New Connection'' strategy. In this strategy, the software considers a much larger number of potential nodes, resulting in greater output variability. Therefore, when using this strategy, we recommend conducting multiple analyses with different weighted degree filters to identify the nodes that most frequently occur in the output. Additionally, the weighted degree filter helps prevent the brand from connecting to peripheral nodes or nodes with a degree that is very low. This allows for selecting a set of nodes that not only improves the brand's connectivity but is also sufficiently central in the discourse of companies' stakeholders.

We point out that the semantic network was created using conference call texts of all companies in the dataset. Those companies encompass various industries, from finance and insurance to transportation and storage or human health. This heterogeneity could have led to varying results in the BNB output when implementing the new connections strategy. However, throughout 30 simulations, only 18 different nodes were identified in the output.

Comparing the output of the new connection strategy and the change weight strategy, we highlight higher output heterogeneity. This is plausible as the change weight strategy focuses on a limited number of nodes, specifically the neighbors of the target node. On the other hand, the new connection strategy considers all nodes in the semantic network. Despite numerous simulations, only three nodes appear more than a third of the time.[5]

Specifically, these three nodes are "anemia" (present in 10 simulations), "FDA" (present in 12 simulations), and "barrel" (present in 21 simulations), while the average frequency of all the nodes appearing in the output is only 4.8. We interpret the output of the BNB as follows.

First, in the years following our analysis period, "anemia" has been shown to be one of the triggering factors leading to the death of COVID-19 patients (Abu-Ismail et al., 2023). Similarly, the Food and Drug Administration (FDA) has arguably been one of the most central institutions in approving diagnostic tests, treatments, and vaccines against COVID-19. Overall, the presence of these two words in the software output could indicate the pandemic's impact on the healthcare system, as well as the crucial intervention later adopted by the FDA, and not least, one of the main causes of death related to COVID-19. With this case study, we do not aim to demonstrate that the BNB software could have identified the causes of COVID-19 deaths two years in advance. On the contrary, our goal is to show that among the over 2000 possible nodes that could have appeared in the output, the most frequent were two words that reflected prominent terms in the subsequent health crisis.

---

[5] Similar strategies might include discussing the nodes that appear in the top quartile or the most frequent nodes. Obviously, it is difficult to give an exact guideline on the pragmatic selection of nodes. Still, we suggest running several simulations and then focusing on the most recurrent output nodes.

Another significant term alongside "FDA" and "anemia" is "barrel", which typically refers to crude oil. Its presence among the top nodes may reflect the concerns experienced by the energy sector during the pandemic. The pandemic caused a drastic reduction in oil demand due to lockdowns and travel restrictions, leading to an oversupply and exceptionally low oil prices. Precisely, oil prices hit rock bottom at the beginning of April 2020, while our analysis was conducted during March. Interestingly, the rebound in oil prices started in the subsequent months. Similarly to "anemia" and "FDA", in this case, the BNB output seems to even more strongly reflect one of the main drivers that put COVID at the center of the discourse in subsequent months.

Although our goal is not to demonstrate the predictive value of the software, overall, the nodes outlined by this case study have then reflected some of the key points that have allowed the COVID term to be central in the discourse. Future research could explore the predictive power of software, especially when the "New Connection" strategy is used.

As previously mentioned, by analyzing conference calls, we can focus our analysis on individual transcripts to better understand the companies most closely related to the BNB output. For instance, after reviewing all the transcripts, we can pinpoint the one with the highest mentions for a specific term among those suggested by BNB.

The transcript with the most frequent mentions of the "barrel" term belongs to VAALCO company, where it is referenced 53 times – which is more than double the number of mentions in the second most referenced transcript, where the term appears only 25 times. Given the significant presence of the "barrel" term in VAALCO's transcript, we decided to further analyze VAALCO's market returns in the weeks following the conference call.

Fig. 3 shows the cumulative returns of one dollar invested in VAALCO and one dollar invested in the industry to which VAALCO belongs from the beginning of the year until June 2020. The stock prices were sourced from the Center for Research in Security Prices (CSRP).

As of January 1, both portfolios exhibited negative returns, a trend prevalent across all industries in the energy sector in the early 2020 s due to a significant drop in oil prices. However, the dynamics changed after VAALCO's conference call on March 10. Following a price spike on March 10 (indicated in Fig. 3 by the first dashed vertical line), VAALCO's returns experienced a sharp decline, negatively surpassing that of the benchmark industry. VAALCO's share prices plummeted from 1.69 per share to a low of 0.71 on March 23, marking a nearly 60 percent drop in less than two weeks. Of greater significance is the trend in prices. While industry-wide prices gradually increased, VAALCO remained stagnant at very low levels. This may suggest that investors may have perceived VAALCO as particularly vulnerable and exposed to the pandemic crisis, at least until the second conference call of the year (indicated by the second vertical dashed line).

This idea is reinforced by the significant presence of COVID-19 in the conference call held by VAALCO the following quarter (on May 12, marked by the second vertical dashed line in Fig. 3). The transcript of March 10 does not mention any pandemic-related terms, but this changes in the subsequent months. On May 12, 2020, the term "Covid" was mentioned six times during the presentation, along with three mentions of the term "pandemic." Furthermore, during the Q&A section, "Covid" is mentioned four more times, along with three mentions of "pandemic".

Lastly, by analyzing the semantic network of the conference calls conducted in May, we discovered that 33 connections were established between COVID-19 and "barrel" and 90 connections between COVID-19 and "FDA".

Once again, our goal is not to demonstrate a predictive power of the software. However, this case study opens the possibility that the software could be useful for identifying key points that regulators should monitor in advance. The fact that connections between COVID-19 and FDA, which were not present in the network analyzed in March, emerged in the following months suggests that these terms contributed to increasing the centrality of the COVID-19 term in the discourse of subsequent months. Similarly, the fact that the VAALCO company was significantly affected by the oil crisis suggests that the BNB software could have been used to monitor this exposure in advance.

## 5. Conclusions and future research directions

Our research focused on the problem of designing effective strategies to improve the connectivity of a target brand (or term). From the algorithmic side, this required tackling a computationally hard problem known as Maximum Betweenness Improvement (MBI) in the literature. Known solutions for this problem (Bergamini et al., 2018; D'Angelo et al., 2016) do not consider realistic constraints (i.e., adversarial and forbidden nodes, constrained budgets, and strengthening existing connections).

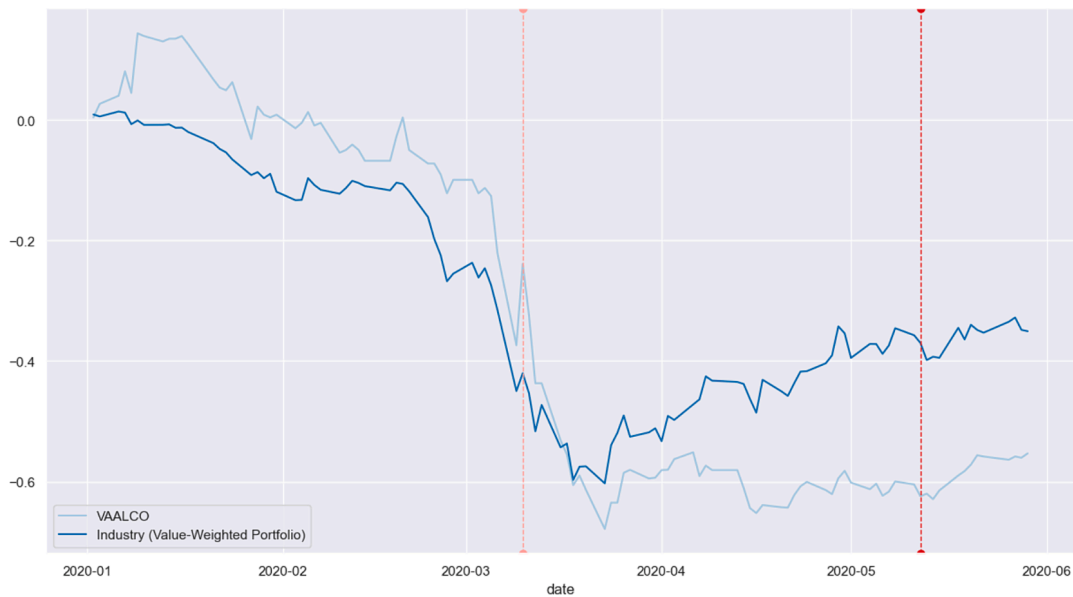

**Fig. 3.** Cumulative abnormal returns of VAALCO compared the energy industry.

This research tries to fill this gap. We developed software designed to enhance brand connectivity within a semantic network. The software is based upon an extended version of the MBI problem that better fits the considered application scenario. The BNB software aims to simulate real-world settings, using budget constraints and accounting for adversarial and forbidden nodes. With its graphical interface, users can analyze the position of one or more brands within the network, identify associations around the brand, and simulate the effect of changes in weights or adding new connections. More generally, the adoption of network visualization tools makes it possible to gain an insight into the network structure and how it affects the connectivity of the nodes of interest.

As also supported by the case studies, the software and the new algorithm offer multiple practical applications. From a managerial perspective, the software can be leveraged to get real-time, data-driven insights into a brand's connectivity within semantic networks. By identifying strategic associations and optimizing discourse pathways, the software can support managers in enhancing brand visibility and positioning and responding swiftly to changes in the brand discourse. Additionally, the user-friendly interface allows for easy customization and simulation of different scenarios, making it a powerful tool for informed decision-making and strategic planning. This ensures managers can effectively leverage big data to maintain a competitive edge and drive brand growth. Finally, the system we presented is not only beneficial for scholars interested in extending the MBI problem but also for managers and practitioners interested in improving the connectivity dimension of their brand. Our tool can support the strategic decision-making process of marketing and communication departments, for example, providing value-added information for better planning of marketing and advertising campaigns.

Concerning the limitations of our approach, while the proposed implementation is fast enough to be effectively applied to real-world instances, large budgets might lead to long running times. Future research could consider improved heuristics to speed up the running times further. The datasets presented in this paper are only intended to illustrate the software's use and suggest potential applications. They are not meant to represent exhaustive case studies, as they are limited in scope and time span. For example, a graph for fashion brands might vary depending on the time of data collection and the launch of specific initiatives for different brands (such as a runway show for the presentation of a new haute couture collection or the sponsorship of a music event).

Concerning the user interface, we exploited a general-purpose visualization tool for large-scale graphs, which runs entirely in the browser and avoids visual clutter and overplotting. On the other hand, it would be interesting to investigate more ad-hoc visualizations for semantic networks that could better highlight critical paths and flows. Furthermore, it would be valuable to evaluate the applicability of our approach beyond semantic networks. This could involve exploring its utility in other scenarios where enhancing the centrality of a node is crucial, such as collaborative innovation networks (Allen et al., 2016).

Lastly, the case studies hint that the software might help make forecasts or anticipate future events. While a dedicated validation study is beyond the scope of this paper, exploring this potential could significantly enhance the software's value for regulators and managers, allowing for more proactive and informed decision-making.

## CRediT authorship contribution statement

**Jacopo Cancellieri:** Writing – review & editing, Writing – original draft, Visualization, Validation, Software. **Walter Didimo:** Writing – review & editing, Writing – original draft, Visualization, Methodology, Conceptualization. **Andrea Fronzetti Colladon:** Writing – review & editing, Writing – original draft, Visualization, Methodology, Conceptualization. **Fabrizio Montecchiani:** Writing – review & editing, Writing – original draft, Visualization, Methodology, Conceptualization. **Roberto Vestrelli:** Data curation, Validation, Visualization, Writing – original draft, Writing – review & editing.

## Declaration of Competing Interest

The authors declare that they have no known competing financial interests or personal relationships that could have appeared to influence the work reported in this paper.

## Data availability

The authors do not have permission to share the online news articles and other textual data. However, they will provide the semantic networks upon reasonable request to the last author.

## Acknowledgments

This work was partially supported by the University of Perugia – Department of Engineering through the program "Fondo Ricerca di Base 2022", project n. RICBA22AFC ("Data analytics e business intelligence per il supporto alle decisioni manageriali nell'era dell'Industria 4.0"). Research also partially supported by: (i) University of Perugia, Fondo Ricerca di Ateneo 2021, Proj. "AIDMIX-Artificial Intelligence for Decision making: Methods for Interpretability and eXplainability"; (ii) MUR PRIN Proj. 2022TS4Y3N - "EXPAND: scalable algorithms for EXPloratory Analyses of heterogeneous and dynamic Networked Data"; (iii) MUR PRIN Proj. 2022ME9Z78 - "NextGRAAL: Next-generation algorithms for constrained GRAph visuALization". The funders had no role in study design, data analysis, decision to publish, or preparation of the manuscript.

## Appendix

Our algorithmic framework was built on top of Graph Tool (Peixoto, 2014), an efficient Python module for manipulating and statistically analyzing graphs. Graph Tool has been extended by adding three algorithms: the first to obtain the data structures to be used in the dynamic computation of betweenness centrality, the second to perform the dynamic computation, and the third for computing the maximum betweenness improvement.

To assess the efficiency of our algorithms, we performed an experimental analysis in which we recorded the running time taken on different inputs and with different tunings (i.e., different values of the parameters). Concerning the hardware and software environment, we tested the algorithms on a server with an Intel Xeon Silver 4214R, 64 G.B. of RAM, the O.S. Ubuntu 20.04.3 LTS, and the 2.43 version of the GraphTool library (Peixoto, 2014). We used three co-occurrence networks of different sizes, and we extracted the largest connected component from each of them, called NET-1, NET-2, and NET-3, in the following. The largest connected component of NET-1 has 1,715 nodes and 11,747 edges, the largest connected component of NET-2 has 8,937 nodes and 164,254 edges, and the largest connected component of NET-3 has 17,307 nodes and 412,912 edges. For each network, we chose four target nodes according to their initial betweenness centrality: a node with the minimum value (0), a node with a low value of about 1,000, a node with a medium value of about 100,000, and a node with the maximum value.

For the sake of presentation, we restricted the parameters of the algorithm as follows: (a) We considered two possible values of budget, namely 10 and 100; (b) we tested both the option in which the algorithm can only increase the weight of existing edges and the option in which the algorithm can also add new edges to the network; (c) the algorithm can use the budget on at most one edge (either an existing edge or a new one, based on the previous option); and (d) we considered both the case in which the binary search looks for a minimum amount of budget that maximizes the improvement and the simpler case in which the whole budget is used.

The results are summarized in Tables A1 to A6. We remark that the execution time of each test is averaged over three repetitions. We observe that the algorithm takes a few seconds on NET-1, at most 2 min on NET-2, and at most 10 min on NET-3. In general, the larger the betweenness of the brand node, the higher the running time. This trend is stronger when the algorithm can only change the weight of existing edges in the network (CHANGE WEIGHT column in the tables). Arguably, this is because central nodes have a large degree, which has an impact on the number of iterations of the algorithm when running with this option. To summarize, the algorithm's running time is reasonable also on a relatively large network. The option that allows the algorithm to add new edges is faster when starting with nodes already having high betweenness; if this is not the case, the option that only allows changing the weight of existing edges appears to be faster.

**Table A1**

Running times of the experiments on NET-1 without the binary search. The column NEW CONNECTION corresponds to the algorithm's option in which new edges can be added to the network, while the column CHANGE WEIGHT corresponds to the option in which the algorithm can only increase the weight of existing edges. The rows MIN, LOW, MEDIUM, MAX correspond to the node with the input network's minimum, low, medium, and maximum value of betweenness.

| | Runtime (seconds) | | | |
| | New connection | | Change weight | |
| Btw Value | Budget-10 | Budget-100 | Budget-10 | Budget-100 |
|---|---|---|---|---|
| Min | 0.615 | 0.945 | 0.333 | 0.329 |
| Low | 0.850 | 1.554 | 0.349 | 0.385 |
| Medium | 1.092 | 1.776 | 2.304 | 6.524 |
| Max | 1.678 | 2.207 | 5.530 | 10.944 |

**Table A2**

Running times of the experiments on NET-1 with the binary search. The column NEW CONNECTION corresponds to the algorithm's option in which new edges can be added to the network, while the column CHANGE WEIGHT corresponds to the option in which the algorithm can only increase the weight of existing edges. The rows MIN, LOW, MEDIUM, MAX correspond to the node with the input network's minimum, low, medium, and maximum value of betweenness.

| | Runtime (seconds) | | | |
| | New connection | | Change weight | |
| Btw Value | Budget-10 | Budget-100 | Budget-10 | Budget-100 |
|---|---|---|---|---|
| Min | 1.010 | 3.923 | 0.327 | 0.380 |
| Low | 1.718 | 8.393 | 0.465 | 0.775 |
| Medium | 1.223 | 4.102 | 3.396 | 27.535 |
| Max | 1.643 | 2.260 | 5.605 | 22.884 |

**Table A3**

Running times of the experiments on NET-2 without the binary search. The column NEW CONNECTION corresponds to the algorithm's option in which new edges can be added to the network, while the column CHANGE WEIGHT corresponds to the option in which the algorithm can only increase the weight of existing edges. The rows MIN, LOW, MEDIUM, MAX correspond to the node with the input network's minimum, low, medium, and maximum value of betweenness.

| | Runtime (seconds) | | | |
| | New connection | | Change weight | |
| Btw Value | Budget-10 | Budget-100 | Budget-10 | Budget-100 |
|---|---|---|---|---|
| Min | 25.803 | 36.451 | 10.185 | 9.119 |
| Low | 47.663 | 65.143 | 10.686 | 10.639 |
| Medium | 45.139 | 66.333 | 14.393 | 24.420 |
| Max | 67.529 | 95.816 | 92.536 | 142.638 |

**Table A4**

Running times of the experiments on NET-2 with the binary search. The column NEW CONNECTION corresponds to the algorithm's option in which new edges can be added to the network, while the column CHANGE WEIGHT corresponds to the option in which the algorithm can only increase the weight of existing edges. The rows MIN, LOW, MEDIUM, MAX correspond to the node with the input network's minimum, low, medium, and maximum value of betweenness.

| | Runtime (seconds) | | | |
| | New connection | | Change weight | |
| Btw Value | Budget-10 | Budget-100 | Budget-10 | Budget-100 |
|---|---|---|---|---|
| Min | 42.365 | 176.859 | 10.015 | 9.361 |

**Table A4** (*continued*)

| Btw Value | Runtime (seconds) | | | |
| | New connection | | Change weight | |
| | Budget-10 | Budget-100 | Budget-10 | Budget-100 |
| --- | --- | --- | --- | --- |
| Low | 111.586 | 330.261 | 11.445 | 16.397 |
| Medium | 53.521 | 171.662 | 15.530 | 72.843 |
| Max | 67.878 | 116.885 | 92.216 | 139.515 |

**Table A5**
Running times of the experiments on NET-3 without the binary search. The column NEW CONNECTION corresponds to the algorithm's option in which new edges can be added to the network, while the column CHANGE WEIGHT corresponds to the option in which the algorithm can only increase the weight of existing edges. The rows MIN, LOW, MEDIUM, MAX correspond to the node with the input network's minimum, low, medium, and maximum value of betweenness.

| Btw Value | Runtime (seconds) | | | |
| | New connection | | Change weight | |
| | Budget-10 | Budget-100 | Budget-10 | Budget-100 |
| --- | --- | --- | --- | --- |
| Min | 145.053 | 234.436 | 48.981 | 54.459 |
| Low | 233.691 | 314.224 | 51.187 | 67.631 |
| Medium | 238.712 | 317.760 | 54.468 | 85.420 |
| Max | 378.278 | 653.807 | 588.642 | 690.569 |

**Table A6**
Running times of the experiments on NET-3 with the binary search. The column NEW CONNECTION corresponds to the algorithm's option in which new edges can be added to the network, while the column CHANGE WEIGHT corresponds to the option in which the algorithm can only increase the weight of existing edges. The rows MIN, LOW, MEDIUM, MAX correspond to the node with the input network's minimum, low, medium, and maximum value of betweenness.

| Btw Value | Runtime (seconds) | | | |
| | New connection | | Change weight | |
| | Budget-10 | Budget-100 | Budget-10 | Budget-100 |
| --- | --- | --- | --- | --- |
| Min | 242.897 | 1136.414 | 49.651 | 57.128 |
| Low | 537.617 | 1579.678 | 56.183 | 118.275 |
| Medium | 532.684 | 1529.157 | 64.129 | 199.055 |
| Max | 379.873 | 638.409 | 589.181 | 689.317 |